# Does Life's Rapid Appearance Imply a Martian Origin?

P.C.W. DAVIES

Australian Centre for Astrobiology, Macquarie University, Sydney, New South Wales, Australia

## ABSTRACT

**The hypothesis that life's rapid appearance on Earth justifies the belief that life is widespread in the universe has been investigated mathematically by Lineweaver and Davis (*Astrobiology* 2, 293–304). However, a rapid appearance could also be interpreted as evidence for a non-terrestrial origin. I attempt to quantify the relative probabilities for a non-indigenous versus indigenous origin, on the assumption that biogenesis involves one or more highly improbable steps, using a generalization of Carter's well-known observer-selection argument. The analysis is specifically applied to a Martian origin of life, with subsequent transfer to Earth within impact ejecta. My main result is that the relatively greater probability of a Martian origin rises sharply as a function of the number of difficult steps involved in biogenesis. The actual numerical factor depends on what is assumed about conditions on early Mars, but for a wide range of assumptions a Martian origin of life is decisively favored. By contrast, an extra-solar origin seems unlikely using the same analysis. These results complement those of Lineweaver and Davis.**

THERE IS A WIDESPREAD belief among astrobiologists that life established itself on Earth with remarkable rapidity once conditions became hospitable. Fossil evidence suggests that life was well advanced by 3.5 Gyr ago, and may have been extant before 3.8 Gyr (Mojsziz *et. al.,* 1996), although direct evidence remains controversial. On the other hand, heavy cosmic bombardment with potentially worldwide sterilizing capability continued until about 3.8 Gyr ago (Maher and Stevenson, 1988; Sleep *et. al.,* 1989). Thus the "window" in which life established itself once conditions became favorable may have been narrow compared with typical geological and astronomical time scales, and also compared with the duration over which Earth has maintained biophilic conditions (~$4 \times 10^9$ years).

The foregoing standard scenario—that life established itself rapidly on Earth following an extended hostile period—is certainly open to question in the light of sparse and contentious data concerning the early Solar System. However, in this paper I do not intend to review that debate. Instead, I wish to address the following question: *If* life appeared rapidly on Earth toward the end of the late heavy bombardment, can any statistical conclusions be drawn from this fact? In an earlier paper, Lineweaver and Davis (2002) asked precisely this question, and were able to derive some interesting conclusions about the likelihood of life emerging on other earth-like planets. This paper is an extension of their analysis.

Let me start by making a general point about the *a priori* probability of biogenesis. There is no known reason why the physical factors that determine the probability $p_L$ of life originating on a given earth-like planet after, say, several billion years, should be correlated with the number of such planets within the galaxy, $N$. The former quantity has to do with organic and physical chemistry, the latter with gravitation, nuclear physics, the scale of the primordial density fluctuations in the early universe, etc. There is therefore no reason to expect, as is sometimes suggested, that $p_L N \sim 1$ (i.e., that there may be a handful of planets in the galaxy with life). In the absence of any linkage between the processes that determine these two parameters, we can reasonably assume that either $p_L N \gg 1$ or $p_L N \ll 1$. An equivalent way of expressing this point is in terms of the expectation time $\tau_L$ for life to emerge *de novo* on an earth-like planet. If $\tau_{planet}$ is a typical astronomical/geological time scale (i.e., billions of years), then $\tau_L \sim \tau_{planet}$ would be an unlikely coincidence, so that either:

$$\tau_L \ll \tau_{planet}$$

or

$$\tau_L \gg \tau_{planet}$$

Scientists have traditionally divided into two broad camps on the matter of how likely life is to emerge given earth-like conditions. The first adheres to hypothesis A: Life is relatively easy to get going from non-life (see, for example, de Duve, 1995). The second prefers hypothesis B: Life is an exceedingly lucky fluke, a statistical accident so improbable that it may have happened only once in the observable universe (see, for example, Monod, 1971). In the former case the galaxy will be teeming with life. In the latter case biogenesis may be very rare or even unique in the observable universe, although life could still exist in more than one location because of a dissemination mechanism such as panspermia.

Hypothesis A accords naturally with the (admittedly fragmentary) facts concerning life's rapid appearance on Earth. If life starts easily and quickly we would expect to see evidence for it soon after conditions on Earth settled down. ~~Given hypothesis A, and~~ Armed with an estimate of *how* quickly life appeared on Earth, it is then possible to make meaningful predictions about the likely value of $p_L$, and hence, if Earth is typical, about the likelihood of life on other earth-like planets (Lineweaver and Davis, 2002).

In this paper, which complements that of Lineweaver and Davis (2002), I consider the conclusions that may be drawn if hypothesis B were correct (i.e., if $\tau_L \gg \tau_{planet}$). This situation was considered by Carter (1983), who used it to explain why, to within a factor of about 2, the time taken for evolution to produce intelligence on Earth is the same as the expected lifetime of the main sequence stage of the sun (alternatively, the duration of habitability of the Earth). Here I shall apply Carter's reasoning to the origin of life, rather than to its evolution.

On the face of it, hypothesis B does not accord well with assumption that life emerged rapidly on Earth. However, scientists who nevertheless believe life is very hard to get going may offer the following alternative explanation: Life did not arise on Earth *de novo,* but came from elsewhere in the universe. If the early Earth was subjected to an ongoing influx of viable organisms, then it would be no surprise if life established itself on Earth soon after conditions permitted. In what follows I examine this alternative explanation by estimating the relative probabilities that terrestrial life (i) originated on Earth, and (ii) came from a nearby earth-like planet. The most promising candidate for (ii) is Mars, although the analysis would apply equally well to Venus or any other body.

**DID TERRESTRIAL LIFE START ON MARS?**

A case can be made that early Mars offered a more favorable environment for biogenesis than Earth (Davies, 1998; Kirschvink and Weiss, 2002), for a variety of reasons. First, the consequences of the cosmic bombardment may have been less severe. Mars avoided the enormous desiccating moon-forming impact that Earth suffered. Moreover, any oceans on Mars would have contained a much lower volume of water than their terrestrial counterparts, thus removing the risk of a global sterilizing heat blanket of rock vapor and superheated steam (Sleep and Zahnle, 1998). Second, being a smaller planet Mars cooled quicker, permitting the early establishment of a deep subsurface zone in which hyperthermophilic organisms could take refuge from the bombardment. Finally, Mars' low surface gravity would have led to the rapid escape of hydrogen from photodissociation of water vapor, thus providing a convenient redox gradient at early times (Kirschvink and Weiss, 2002). Taking all these factors together, it seems that Mars may have been suitable for life as long ago as 4.5 Gyr. If so, then it had a head start over Earth. There is a chance that life may then have spread to Earth cocooned inside rocks ejected from Mars by impacts (Melosh, 1988; Davies, 1995, 1996; Weiss *et.al.,* 2000), a phenomenon that I call transpermia, to distinguish it from the older idea of panspermia. How can we quantify the relative plausibility of this scenario compared with that of a terrestrial origin of life?

Suppose that, in accordance with hypothesis B, life is very hard to start (i.e., that the expectation time for life to emerge spontaneously on a suitable Earth-like planet is very much longer than the habitability duration of that planet). Following Carter (1983), I assume that the emergence of life involved a sequence of *n* independent highly improbable random processes (e.g., the formation of a particular key molecule or set of molecules in one location.) The probability that *one* such process is successfully completed after a time *t* is then:

$$p(t) = (1 - e^{-t/\tau}) \qquad (1)$$

where $\tau$ is approximately the expectation time for the process to occur. The probability for *n* such processes to be completed, if they are statistically independent, is then:

$$p(t) = (1 - e^{-t/\tau})^n \qquad (2)$$

where for simplicity I am assuming the processes are equally (im)probable.

I now consider the relative sizes of the windows of opportunity for Mars and Earth to generate life. Define the *biogenesis window* as the length of time on a given planet over which conditions permit biogenesis to occur; call this $T_M$ and $T_E$ for Mars and Earth, respectively. Then the relative probabilities $p_M$ and $p_E$ that life will have emerged at some time before the end of that window on those respective planets is given by:

$$p_M/p_E = (1 - e^{-T_M/\tau_M})^n (1 - e^{-T_E/\tau_E})^{-n} \qquad (3)$$

where I allow for the possibility that $\tau$ might differ between the two planets. This would be the case if the conditions on, say, Mars were more conducive to biogenesis than on Earth; that is, if the probability *per unit time* for life to arise on Mars were greater than that on Earth. (See, for example, Kirschvink and Weiss, 2002.)

In accordance with hypothesis B, $T_M, T_E \ll \tau_M, \tau_E$, so Eq. 3 may be approximated:

$$p_M/p_E \cong (T_M \tau_E / T_E \tau_M)^n \qquad (4)$$

Now let me insert some numerical values. Estimates of how broad the window of opportunity was on Earth vary greatly from zero to several hundred million years (Lineweaver and Davis, 2002). If, for example, life formed on or near the Earth's surface, then biogenesis is unlikely until the late heavy bombardment abated around 3.8 Gyr ago. On the other hand, if life formed ~2 km beneath the seabed, offering effective refuge against the bombardment, then the biogenesis window would open as soon as the Earth's crust cooled below, say, 120°C at that depth (about 4 Gyr ago). For the purposes of illustration, I shall take $T_E \sim 10^8$ years, and suppose that life was established on Earth 3.9 Gyr ago. If Mars was indeed suitable for biogenesis from 4.5 Gyr, then $T_M \cong 6 \times 10^8$. From Eq. 4 one infers a relative advantage for Mars of $6^n(\tau_E/\tau_M)^n \cong 10^{0.8n}(\tau_E/\tau_M)^n$.

Set against this advantage is the requirement that life must successfully propagate from Mars to Earth during the $10^8$ year terrestrial "window"—the time after which Earth became habitable and before life is known to have appeared. For simplicity, I assume a fixed probability of viable transfer per unit time, so that the transfer probability $p_{\text{trans}}$ in the window $T_E$ is given by:

$$(T_M/T_E)^n (1 - e^{-T_E/\tau_{\text{trans}}}) \qquad (5)$$

where $\tau_{\text{trans}}$ is of order the expectation time for the successful transfer of life. By hypothesis, Martian organisms arriving before the $10^8$ year window $T_E$ would not have survived long-term. Detailed studies by Mileikowsky *et. al.* (2000) and Weiss *et al.* (2000) suggest quite a high probability of the successful transfer of viable organisms riding in impact ejecta between Mars and Earth. Their work implies a value for $\tau_{\text{trans}} \sim 10^6$ years; at any rate, one may safely assume that $T_E \gg \tau_{\text{trans}}$ so the last factor on the right-hand side of Eq. 5 is ~1.

The foregoing analysis may seriously underestimate the relative advantage for Mars. The reason for this is because it is not sufficient in our analysis for life to merely get started: it also has to survive to the present day. Because we are almost completely ignorant of the nature and setting of the first life forms, it is hard to know with what, in addition to the cosmic bombardment, the first cells had to contend. But it is easy to think of many hazards that would threaten to snuff out any nascent life. Here are a few:

1. The physical and chemical environment most suited to generating life from nonlife may have been very different from the environment in which life multiplies and flourishes. For example, hot, dry conditions might be needed to make certain key organic molecules, but cool, wet conditions favor microbe replication. In the absence of a convenient juxtaposition of the two environments, life may get started but never have a chance to progress.
2. To advance from a toehold to establishing itself in a secure manner, life must spread as widely as possible to survive the environmental insults that follow. It must also mutate and adapt fast enough to keep pace with changing conditions. There will be a critical bottleneck for a newly formed microbial colony to break out of its niche and percolate through the wider environment. This niche is likely to be very atypical of the whole environment, and reaching similar niches may be very hard.
3. Modern microbial life has evolved many sophisticated mechanisms (e.g., the formation of spores) to withstand environmental insults and thus increase the chances of surviving transit between niches. But early life would have lacked this robustness, making it vulnerable to sudden environmental changes.
4. Early life was almost certainly chemotrophic. If life began in a suitable but isolated chemical reservoir, it would rapidly multiply until all resources had been exhausted, and then stop. For life to spread and evolve, there must be a continual throughput of the necessary materials and free energy for metabolism. Only in special geological settings would these conditions be met.

A consideration of these and other requirements that would need to be met before indefinite survival of newly formed life is assured suggests that life would have to form many times over before the "experiment" succeeded. If there were *m* successive and independent episodes of biogenesis before life became well enough established to survive for ~1 Gyr, then the Martian advantage factor (Eq. 4) becomes:

$$p_M/p_E \cong (T_M \tau_E/T_E \tau_M)^{nm} \cong 10^{0.8nm}(\tau_E/\tau_M)^{nm} \quad (6)$$

The numerical factor favoring Mars is certainly significant. If just a single highly improbable step is needed for life to form, and if nature's only "experiment" with life survives all subsequent insults, then the probability of biogenesis on Mars at some time prior to 3.9 Gyr is (given my assumptions) about six times higher than on Earth, if we assume equal prior probabilities per unit time for biogenesis (i.e., $\tau_E = \tau_M$).

Naturally one can question the foregoing illustrative numbers, but the key point about the result (Eq. 6) is this. The "Martian advantage" rises sharply if $n,m > 1$. (By hypothesis B, $n,m < 1$ is ruled out.) So long as Mars has either a longer window of opportunity than Earth, or a greater prior probability of biogenesis ($\tau_M < \tau_E$), or both—whatever the combined factor is—then the right-hand side of Eq. 6 is >1, and each additional step $n$ or $m$ escalates the probability of a Martian origin for terrestrial life. That is the central result of this paper.

What can be said about the value for $n$? Some biochemists have argued that many crucial improbable steps are needed. If $n = 10$, for example, then $p_M/p_E \cong 10^8$, and the odds favoring a Martian origin of terrestrial life become overwhelming. Regarding $m$, one might expect on *a priori* grounds that many episodes of biogenesis would be required before one group of organisms established an unshakable presence on a planet. If one takes, for example, $m = 100$, and $n = 10$ then (with $\tau_M = \tau_E$) Eq. 6 yields the staggering factor of $10^{800}$!

On the other hand, the argument of Carter (1983) can be invoked to place a bound on the total number of highly improbable steps needed for *intelligent* life to emerge on Earth, which therefore places at least as strong a bound on $nm$. This bound depends on the length of time left in the future during which Earth might remain suitable for the emergence of intelligent life. Solar evolution models suggest a time ~1 Gyr, in which case Carter's formula yields a modest value:

$$nm < 4 \quad (7)$$

It is possible to imagine that, say, $nm = 2$, consistent with the Carter bound, but still leading to a 97.5% probability that terrestrial life started on Mars (excluding other potential biogenesis sites).

The conclusion is that if hypothesis B is correct (and given my various assumptions), and if the expected time for biogenesis to occur on a given planet is substantially greater than typical geological/astronomical time scales, then the likelihood of terrestrial life originating on Mars is significantly greater than the likelihood of an indigenous origin. As stressed, the level of plausibility of a Mars genesis is very sensitive to the number of unlikely steps $n$ involved in biogenesis, and the number of "trials" $m$ expected before nascent life manages to establish itself permanently. Even for plausible and conservative values of $n$ and $m$, the likelihood of a Mars origin can be very high. Thus, scientists who believe that biogenesis involved a series of exceedingly rare accidents should strongly favor a Martian over a terrestrial origin. This conclusion would be valid even assigning equal prior probabilities $\tau_M = \tau_E$. If one also argues that $\tau_M < \tau_E$ (e.g., Kirschvink and Weiss, 2002), then the probability of a Mars origin is correspondingly greater. Note that if $\tau_M < \tau_E$ then the argument carries through even if one drops the assumption of a longer Martian window of opportunity. The "Martian advantage" (Eq. 6) is then $(\tau_E/\tau_M)^{nm}$, which rises sharply with $n$ and $m$. Even a modest environmental advantage for Mars would translate into a high probability for a Martian origin of life if $n$ or $m > 1$.

Support for these conclusions comes from the work of Hanson (1998), who has demonstrated that if life arose from a sequence of easy and hard steps constrained to be completed during a fixed time interval, then the expected time left over after the final step is approximately equal to the time taken by the hardest step. However, account must be taken of the selection effect occasioned by our own existence, which requires an extended period of habitability to allow for the evolution of intelligence. This too may have involved hard steps. But by hypothesis B, I assume that the first step—biogenesis—was the hardest. Hanson's argument then leads to the prediction that the habitability time left on Earth from the present epoch should be approximately the time required for biogenesis to occur. Taking the estimate ~1 Gyr for the former leads to a serious contradiction with the hypothesis that life began on Earth in a $10^8$ year window, but is entirely consistent with a Martian origin over a time interval of ~0.6 Gyr.

## DID LIFE START BEYOND THE SOLAR SYSTEM?

If terrestrial life originated beyond the Solar System, we may consider the possibility of much longer prebiotic durations, $T_*$. Depending on assumptions about heavy element production and planetary system formation, Earth-like planets may have existed as long ago as $10^{10}$ years. In which case:

$$T_*/T_E \sim 10^2 \tag{7}$$

In place of Eq. 5 we now have:

$$(T_*/T_E)^{nm}(1 - e^{-T_E/\tau_{trans}}) \sim 10^{2nm}(1 - e^{-T_E/\tau_{trans}}) \tag{8}$$

The probability for the transfer of viable microorganisms across *interstellar* space is exceedingly low, so Eq. 8 is, to a good approximation:

$$10^{2nm} T_E/\tau_{trans} \tag{9}$$

This expression must be modified by a factor representing the probability that an ancient planetary system passed to within a few light years of the Solar System during the "window" $T_E$ around 4 billion years ago. There is evidence (Napier and Clube, 1997) from the comet bombardment record of a periodicity of 30 Myr, this being the interval between successive passages of the Solar System through the galactic plane, resulting in stellar encounters that disrupt the Oort cloud and generate a spate of impacts. Thus an encounter within the $10^8$ year window $T_E$ seems a reasonable possibility. All in all, then, the modification factor is of order unity, and Eq. 9 remains a fair approximation.

An interstellar origin of terrestrial life becomes a plausible hypothesis if Eq. 9 is ~1, which requires:

$$\tau_{\text{trans}} < 10^{8+2nm} \text{ years} \qquad (10)$$

Melosh (2003) has studied interstellar transpermia, and estimates a probability of $2.5 \times 10^{-5}$ for the impact on Earth of a rock from an Earth-like extrasolar planet during the 4.5 Gyr history of the solar system, or $10^{-14}$ y$^{-1}$. They also estimate an average sojourn time within the ejecting planetary system of 50 Myr prior to ejection by a giant planet, assuming the solar system is typical. This is close to the limit of the survival time against radiation damage for even the most radiation-resilient known organisms, so the probability needs to be reduced by several orders of magnitude when discussing the chances of the viable transfer of organisms. Taking all these factors into consideration, it seems reasonable to assume that $\tau_{\text{trans}}$ for rocky panspermia is $>10^{16}$ years, which gives the interesting inequality:

$$nm > 4 \qquad (11)$$

before an extra-solar origin is favored. This inequality may be compared with Carter's bound on $nm$, adapted to galactic rather than terrestrial circumstances. The habitability window for the galaxy is between one and few stellar lifetimes. At least one generation of stars must be permitted to make carbon, while the paucity of sun-like stars after several stellar generations would drastically reduce the chances of the emergence of intelligent life. So very roughly, intelligent life has emerged between about one-quarter to ½ way through this galactic window of opportunity, leading to a Carter bound of

$$nm < 4 \qquad (12)$$

Clearly Eq. 11 is in conflict with Eq. 12, which suggests that an interstellar origin of terrestrial life via transpermia is *not* a plausible hypothesis.

## CONCLUSION

The problem of life's origin is bedevilled by the fact that we have access to only one sample of life. Because we have selected that sample by our very existence, statistical inferences are hazardous. Conclusions may be drawn only by augmenting the science with a philosophical principle. Two contrasting examples are the principle of mediocrity and the so-called anthropic principle (Barrow and Tipler, 1986). According to the former, the Earth, including its biosphere, is a typical member of a large class of planets. By contrast, the anthropic principle asserts that we have selected an atypical location in the universe by virtue of our existence. The former principle may be used to argue that life is common in the universe (what I have called hypothesis A), the latter that it is rare or even unique (my hypothesis B).

In spite of this diversity of positions, some valid statistical conclusions may be drawn if we invoke an important additional possibility: that life may have appeared on Earth rather rapidly. ~~Lineweaver and Davis (2002) have analyzed the consequences of this under both hypotheses A.~~ In this paper I consider ~~the corresponding analysis for~~

hypothesis B: that life is a freak accident. I argue that if life emerged from a series of highly improbable chemical and physical steps, as is widely assumed by biologists, then a Martian origin for terrestrial life is probable, or even highly probable. The probability of a Martian origin rises sharply with the number of assumed independent improbable steps involved, and with the number of geneses needed before life establishes itself firmly. For only a handful of such steps the probability becomes overwhelming. Taking the results of this paper together with those of Lineweaver and Davis (2002), the conclusion seems to be that either life arises readily and is common throughout the universe, or it is exceedingly rare and probably originated on Mars.

Though these analyses are interesting, caution is necessary, as they are based on many assumptions about biology and astronomy that offer a wide scope for disagreement. In particular, the relative durations of the biogenesis windows for Earth and Mars might depend on a number of unknown or as yet poorly understood factors. Second, the assumption that Earth and Mars enjoyed single uniform windows of opportunity for biogenesis is clearly highly simplistic. The bombardment of Earth by very large impactors was probably sporadic, sparse, and not necessarily completely globally sterilizing. Rather than a single long window, there could have been a succession of shorter windows. Even given a totally sterilizing impact, there remained the possibility that some ejected material could offer a refuge for terrestrial organisms, returning them to Earth at a later date when conditions had settled down. Some ejected terrestrial material would have reached Mars, opening up the possibility of a complex interchange of life back and forth between the two planets. These additional features would complicate the simple calculation given here. Finally, the probabilities calculated here are conditioned on the information that primitive life definitely existed on Earth at 3.5 Gyr or more ago. One could also discuss the relative probabilities that life might arise on Earth or Mars at *any* time subsequent to these planets becoming habitable, subject only to the condition that intelligent life should emerge before Earth ceases to be habitable. This would involve making assumptions about the pace of evolution.

In spite of these complications, the general conclusions offered in this paper are robust, and can be refined as our knowledge and understanding of early Mars and early Earth improves. In the absence of a second sample of life this type of analysis is probably the best hope we have for comparing alternative classes of hypotheses about life's origin and prevalence in the Universe.

## ACKNOWLEDGMENTS

I am grateful to Pauline Davies for drawing my attention to the importance of the "survival factor" *m,* and to Malcolm Walter, Michael Paine and Robin Hanson for critical comments on an early draft of this paper. I have also benefited from helpful discussions with Charles Lineweaver, and from correspondence with Joseph Kirschvink and Norman Sleep.

## REFERENCES

Barrow, J.D. and Tipler, F.J. (1986) *The Anthropic Cosmological Principle,* Clarendon Press, Oxford.
Carter, B. (1983) The anthropic principle and its implications for biological evolution. *Philos. Trans. R. Soc. Lond.* A310, 347–355.

Davies, P.C.W. (1995) *Are We Alone?,* Penguin, London.
Davies, P.C.W. (1998) *The Fifth Miracle: The Search for the Origin of Life*, Allen Lane, Penguin Press, London; (2003) 2nd edit., *The Origin of Life,* Penguin, London.
de Duve, C. (1995) *Vital Dust,* Basic Books, New York.
Hanson, R. (1998) Must early life be easy? The rhythm of major evolutionary transitions. Available on-line at: http://hanson.gmu.edu/hardstep.pdf. See also "The Great Filter" at http://hanson.gmu.edu/greatfilter.html.
Kirschvink, J.L. and Weiss, B.P. (2002) Mars, panspermia, and the origin of life: where did it all begin? *Palaeolont. Electron.* 4(2), 8–15.
Lineweaver, C.H. and Davis, T.M. (2002) Does the rapid appearance of life on Earth suggest that life is common in the Universe? *Astrobiology* 2, 293–304.
Maher, K.A. and Stevenson, D.J. (1988) Impact frustration of the origin of life. *Nature* 331, 612–614.
Melosh, H.J. (1988) The rocky road to Panspermia. *Nature* 332, 687.
Melosh, H.J. (2003) Exchange of meteorites (and life?) between stellar systems. *Astrobiology* 3, 207–215.
Mileikowsky, C., Cucinotta, F., Wilson, J.W., Gladman, B., Horneck, G., Lindgren, L., Melosh., H.J., Rickman, H., Valtonen, M.J., and Zheng, Z.G. (2000) Natural transfer of viable microbes in space. Part 1: From Mars to Earth and Earth to Mars. *Icarus* 145, 391–427.
Monod, J. (1971) *Chance and Necessity,* trans. A. Wainhouse, Collins, London.
Mojzsis, S.J., Arrhenius, G., McKeegan, K.D., Harrison, T.M., Nutman. A.P., and Friend, C.R.L. (1996) Evidence for life on Earth before 3500 million years ago. *Nature* 384, 55–59.
Napier, W.M. and Clube, S.V.M. (1997) Our cometary environment. *Rep. Prog. Phys.* 60, 293–343.
Sleep, N.H. and Zahnle, K. (1998). Refugia from asteroid impacts on early Mars and the early Earth. *J. Geophys. Res.* 103, 529–544.
Sleep, N.H., Zahnle, K.J., Kasting, J.F., and Morowitz, H.J. (1989) Annihilation of ecosystems by large asteroid impacts on the early Earth. *Nature* 342, 139–142.
Weiss, B.P., Kirschvink, J.L., Baudenbacher, F.J., Vali, H., Peters, N.T., Macdonald, F.A., and Wikswo, J.P. (2000) A low temperature transfer of ALH84001 from Mars to Earth. *Science* 290, 791–795.